\documentclass[twocolumn,preprintnumbers    ,nofootinbib,aps,prd,floatfix,superscriptaddress]{revtex4-2}

\usepackage{graphicx}
\usepackage{lmodern}
\usepackage{dcolumn}
\usepackage{bm}

\usepackage[left=1in, right=1in, top=1.2in, bottom=1.2in, includefoot, headheight=13.6pt]{geometry}

\usepackage{amsmath}
\newcommand{\sech}{\mathrm{sech}}
\usepackage{amssymb}
\usepackage{booktabs}
\usepackage[utf8]{inputenc}
\usepackage[T1]{fontenc}
\usepackage{mathptmx}
\usepackage{etoolbox}
\usepackage{float}
\usepackage[separate-uncertainty=true]{siunitx}
\usepackage{graphicx}
\usepackage{subcaption}
\usepackage{caption}
\usepackage{natbib}
\usepackage{ragged2e}
\usepackage{mathtools}
\usepackage{array}
\usepackage[table,xcdraw]{xcolor}
\usepackage{array}
\usepackage{geometry}
\usepackage{colortbl}
\usepackage{braket}

\usepackage{setspace}
\usepackage{hyperref}
\usepackage{comment}
\usepackage[table]{xcolor}
\usepackage{colortbl}

\begin{document}

\title{The Doppler effect of the Milky Way rotation on mHz gravitational waves}

\author{Giorgio Mentasti}
\email{giorgio@apc.in2p3.fr}
\affiliation{Universit\'e Paris Cit\'e, CNRS, Astroparticule et Cosmologie, F-75013 Paris, France}

\author{Quentin Baghi}
\affiliation{Universit\'e Paris Cit\'e, CNRS, Astroparticule et Cosmologie, F-75013 Paris, France}

\begin{abstract}
We model the stellar density and the rotation curve of the Milky Way and compute the
gravitational-wave spectrum of its unresolved binaries as a function of direction. Because the
Galaxy rotates, the waves emitted along each line of sight are Doppler-shifted by a different
amount, and we derive for the first time the resulting observed anisotropic spectrum, which no
longer factorizes into a spectral shape times an angular template. We then quantify the
observability of this distortion with space-based interferometers. With their nominal noise
budgets, neither LISA nor a LISA+Taiji network can detect it: the residual between the models
that include and neglect the rotation has a signal-to-noise ratio
${\rm SNR}_{\rm diff}\simeq0.2$ and $0.7$ after two
years, and the induced bias on the foreground parameters stays below $0.1\sigma$. A tenfold
better noise amplitude, of the order contemplated for post-LISA concepts, raises the residual to
${\rm SNR}_{\rm diff}\simeq4.6$ for the network, where the effect becomes measurable and
displaces the spectral cutoff by half of the statistical error.
\end{abstract}

\maketitle


\section{Introduction}
\begin{figure*}[t]
	\centering
	\includegraphics[width=\textwidth,clip]{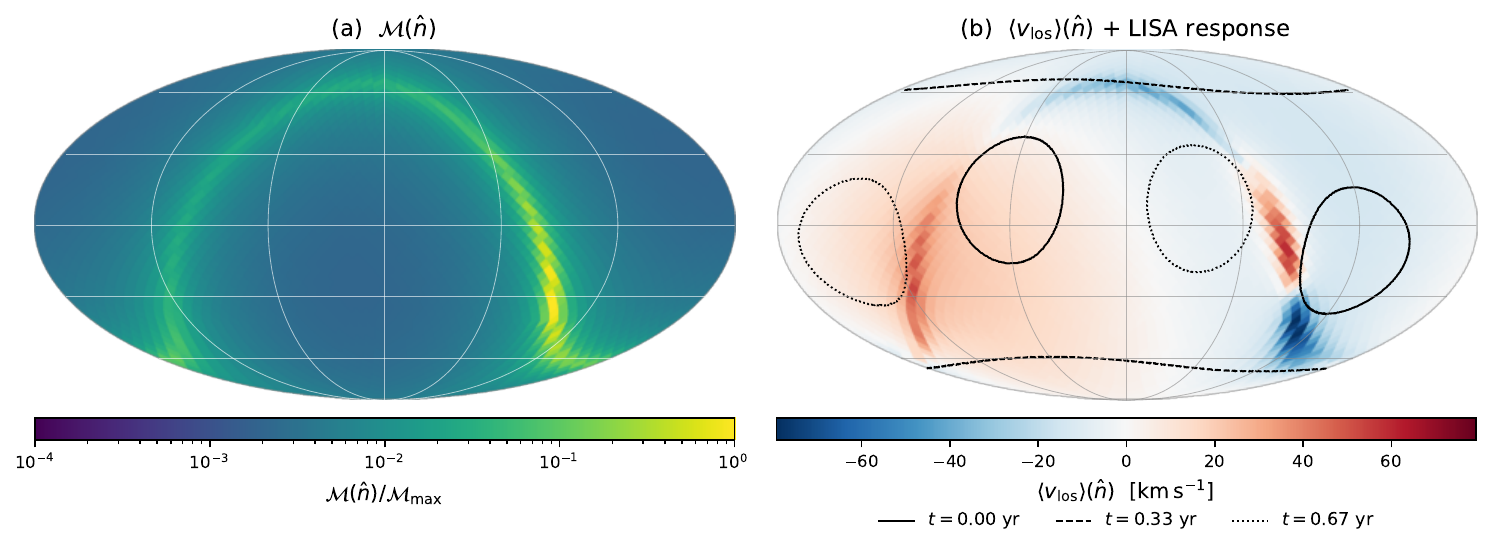}
	\caption{\emph{Left panel:} Mollweide projection, in celestial coordinates, of the angular distribution of the power spectrum of the galactic background $\mathcal{M}(\hat n)$ in Eq.~\eqref{spectrum0}. \emph{Right panel:} density-weighted mean LoS velocity $\langle v_{\rm los}\rangle(\hat n)$ of Eq.~\eqref{eq:vlos}, in the same projection; positive values correspond to receding regions. The black contours show the level curves at $80\%$ of the maximum of the LISA sky response of the $A+E$ channels at $f=3$ mHz, for three epochs separated by a third of a year. All three are the same level curve, drawn at $t=0$, $1/3$ and $2/3$ yr (solid, dashed and dotted respectively, as indicated in the legend). See Fig.~6 of Ref.~\cite{Digman:2022jmp} for a comparable rendering of the LISA antenna pattern in sky projection.}
	\label{fig:map}
\end{figure*}

The upcoming Laser Interferometer Space Antenna (LISA) \cite{Baker:2019nia} will open a new window on the low-frequency gravitational wave (GW) Universe, enabling the detection of stochastic gravitational wave backgrounds~(SGWBs) across a broad range of frequencies. Among the most promising targets for LISA are the astrophysical SGWBs, which arise from the superposition of unresolved sources such as compact binary coalescences, stellar remnants, and other galactic and extragalactic processes \cite{Karnesis:2021tsh,Babak:2023lro,Boileau:2025jkv}. These backgrounds encode valuable information about the astrophysics of their sources, the dynamics of their host environments, and the large-scale structure of the Universe.
An expected strong component of the SGWB is the galactic background, produced by the myriad of unresolved sources within the Milky Way, mainly white dwarfs. Unlike cosmological SGWBs, whose detection remains uncertain and dependent on the details of early-Universe physics, the galactic SGWB is an almost certain signal for LISA. Its detection and characterization will provide a unique opportunity to probe the astrophysics of galactic binaries~(GBs) and other compact objects, complementing the information extracted from individually resolved sources.
However, distinguishing the galactic SGWB from instrumental noise and other astrophysical foregrounds is a formidable challenge. One of the most promising strategies to achieve this separation is to exploit the anisotropies imprinted in the signal by the non-uniform distribution of sources within the Galaxy\cite{boileau2021b,Mentasti:2023uyi,hindmarsh2024,Pozzoli:2024wfe,criswell2024,Buscicchio:2025zeb}. The fact that the anisotropy of the galactic foreground, combined with the motion of the LISA constellation, produces a characteristic annual modulation of the measured power has been recognized since the early studies of the LISA confusion noise~\cite{Seto:2004ji,Timpano:2005gm,Edlund:2005ye}. Its consequences for the detectability of individual sources in a time-varying background have been quantified more recently in Ref.~\cite{Digman:2022jmp}.

In this work, we focus on a previously overlooked effect: the rotation of the Galaxy. The motion of stars in the Milky Way induces a Doppler shift in the GWs they emit, which varies as a function of direction on the sky and therefore breaks the factorization of the foreground into a spectral shape times an angular template. Whether this matters in practice is a quantitative question, and a pressing one. Together with resolved sources, the galactic SGWB provides a comprehensive picture of the binary population in the Galaxy, and it plays a key role in constraining the astrophysics of GBs. A biased measurement of its parameters would therefore lead to incorrect astrophysical interpretations, such as misestimating the abundance, mass distribution, or evolutionary history of GBs. 
We compute the effect exactly within our Galaxy model and find that, for the nominal LISA noise budget, it is well below the detection threshold and induces parameter biases below a tenth of the statistical uncertainty. We then quantify what would be required for it to matter, considering a LISA+Taiji network and instruments with a lower noise floor, and we derive the intrinsic ceiling that no reduction of the instrumental noise can exceed.

To model this effect, we adopt a detailed and physically motivated description of the Galaxy. The frequency spectrum of the SGWB is informed by simulations \cite{Karnesis:2021tsh}, while its angular distribution is derived from a bulge+disk model of the stellar density \cite{Nelemans:2003ha,PhysRevD.86.124032}, integrated along the line of sight.
The kinematics of the stars are described using a smooth fit to the Galactic rotation curve, based on observational data \cite{McGaugh:2018ruf}. Crucially, we derive the correct formula for the Doppler boost induced by the Galaxy's rotation, which, to our knowledge, has never been computed before in the literature. Unlike previous approaches that assumed a single boost direction for the entire background \cite{Heisenberg:2024var,Fumagalli:2026naa}, our model accounts for the fact that each line of sight has a different relative velocity with respect to the Sun, depending on the location and motion of the sources along that direction.

The paper is organized as follows. In Section \ref{sec:spectrum}, we derive the GW spectrum in the rest frame and observer frame, including the Doppler boost from the Galaxy's rotation, and we discuss in detail the assumptions that enter this construction. In Section \ref{sec:forecast}, we present our forecast methodology, based on a Fisher matrix formalism supported by Markov chain Monte Carlo (MCMC) sampling, to assess the detectability of the effect and the biases introduced by neglecting it. Section \ref{sec:implications} discusses the consequences for existing foreground-characterization pipelines and compares the Doppler shift with other kinematic effects of similar physical origin. We conclude in Section \ref{sec:conclusions} with a discussion of the implications of our findings for LISA and future GW observatories. Throughout, we consider both LISA alone and the LISA+Taiji network, and we quantify how the observability of the effect scales with the instrumental noise level. The technical details of the model we adopt for the Galaxy --- its density profile, the source-frame spectral shape, the geometry and the parametrization of the stellar velocities --- are collected in Appendix \ref{sec:galaxy}, to which the reader is referred for the definitions used throughout. The derivation of the bound on the parameter bias used in Section \ref{sec:results} is given in Appendix \ref{sec:bound}.

\section{Spectrum computation}
\label{sec:spectrum}
We define the observed GW power spectrum as a function of the observed frequencies $f$ and sky angles $\hat n$ as
\begin{align}
\label{eq:power-spectrum}
\langle h_\lambda(f,\hat n)h_{\lambda'}(f',\hat n')\rangle=\delta_{\lambda\lambda'}\delta(f-f')\delta^{(2)}(\hat n-\hat n')S_h(f,\hat n)\,,
\end{align}
where $h_\lambda(f,\hat n)$ comes from the superposition of all the waves emitted by sources on the line of sight at $\hat n$, and $\langle\rangle$ designates the ensemble average. Eq.~\eqref{eq:power-spectrum} assumes that sources at different frequencies and sky locations are emitted by different galactic binaries and therefore are completely uncorrelated. To compute an explicit form of this observed spectrum, let's assume that the GW energy emitted by a portion of matter of the Galaxy is proportional to the number of sources present there, and therefore proportional to the density $\rho$. Then let's call $dr$ the line element at the line of sight $\hat n$.
A volume element at distance $r$, located at coordinates $\vec x(r)$ will source a spectrum \textit{in its rest frame} following the spectral shape $S_h^G(f)$, defined in Eq.\eqref{eq:PSD_galaxy}, and decreasing quadratically in the distance
\begin{align}\label{eq:Shrf}
S_h^{\rm rf}(f,\hat n)=I_0 \left(\frac{r_0}{r}\right)^2 S_h^G(f)\,,
\end{align}
while in the rotating frame
\begin{align}
S_h^{v}(f,\hat n)=I_0 \left(\frac{r_0}{r}\right)^2\frac{S_h^G(\mathcal{D}(-\hat n,\vec v(\vec x))f)}{\mathcal{D}(-\hat n,\vec v(\vec x))}\,,
\end{align}
where $\vec v(\vec x)$ is the relative velocity of the volume element at direction $\vec x$ with respect to the Sun, and $\mathcal{D}(\hat{k}, \vec v)$ is the Doppler coefficient, defined by
\begin{align}\label{eq:doppler}
{\mathcal D}(\hat n,\vec v)&\equiv\frac{1+\hat n\cdot\vec v}{\sqrt{1-\beta^2}}\,,
\end{align}
where $\beta=|\vec v|/c$.
Since the GWs emitted by each source are uncorrelated, then the spectrum at the direction $\hat n$ will be
\begin{align}\label{eq:Sh_full}
S_h(f,\hat n)&=\int_0^\infty dr \,\Sigma(r)\rho(\vec x(r))S_h^v(f,\hat n)\nonumber\\
&=4\pi I_0\,r_0^2\int_0^\infty dr \,\rho(\vec x(r))\frac{S_h^G(\mathcal{D}(-\hat n,\vec v(\vec x(r)))f)}{\mathcal{D}(-\hat n,\vec v(\vec x(r)))}\,,
\end{align}
where $\Sigma(r)=4\pi r^2$ is the area of the sphere at distance $r$.
The spectrum that one obtains by ignoring the peculiar velocities effect is obtained by setting $\vec v(\vec x)=0$ everywhere, therefore having $\mathcal{D}=1$ and
\begin{align}\label{spectrum0}
S_h^0(f,\hat n)= 4\pi I_0\,r_0^2 S_h^G(f)\int_0^\infty dr \,\rho(\vec x(r))\equiv S_h^G(f)\,\mathcal{M}(\hat n)\,,
\end{align}
where $\mathcal{M}(\hat n)$ is the anisotropic map coming from integrating the total matter mass along the line of sight $\hat{n}$, whose plot, in celestial coordinates, is displayed in the left panel of Figure \ref{fig:map}. Note that in this case, the frequency dependence and the anisotropy completely factorize.

The quantity that controls the departure from this factorized form is not $\mathcal{M}(\hat n)$ but the anisotropic Doppler factor $\mathcal{D}$, i.e.\ ultimately the density-weighted mean line-of-sight (LoS) velocity along each direction,
\begin{align}\label{eq:vlos}
\langle v_{\rm los}\rangle(\hat n)\equiv\frac{\int_0^\infty dr\,\rho(\vec x(r))\,\vec v(\vec x(r))\cdot\hat n}{\int_0^\infty dr\,\rho(\vec x(r))}\,,
\end{align}
with the convention that $\hat n$ points from the observer towards the source, so that a positive $\langle v_{\rm los}\rangle$ corresponds to a receding region and hence to a redshift of the observed spectrum. This map is shown in the right panel of Figure \ref{fig:map}. It spans from $-83$ to $+62$ km/s, i.e.\ $|v_{\rm los}|/c\lesssim 2.8\times 10^{-4}$: the half of the sky towards which the local standard of rest moves is blueshifted and the opposite half is redshifted, with sharp features along the Galactic plane and a compact quadrupolar pattern around the bulge, where lines of sight pass close to the Galactic centre and pick up large velocity gradients.

Two features of Figure~\ref{fig:map} explain why the effect neither vanishes nor is trivially degenerate with a rescaling of the amplitude. First, the density weighting is essential: the sky average of $\langle v_{\rm los}\rangle$ weighted by $\mathcal{M}(\hat n)$ is only $\simeq-4.8$ km/s, more than an order of magnitude smaller than the peak values, so that most of the Doppler shift cancels when the sky is observed isotropically. Second, LISA does not observe the sky isotropically: its response is quadrupolar and rotates over the year, as illustrated by the contours superimposed on the right panel of Figure~\ref{fig:map}. The detector therefore weights the red- and blue-shifted lobes unequally, and by a different amount at each epoch. It is precisely this incomplete cancellation, modulated annually, that produces the residual quantified in Section \ref{sec:forecast}, and it is also the reason why that residual is so much smaller than the local Doppler shift.

\subsection{The assumption of a universal source-frame spectral shape}
\label{sec:assumption}

The shape of $S_h^G$ of Eq. \eqref{eq:Shrf} is controlled by the distribution of orbital frequencies of the unresolved binaries. Its low-frequency behaviour, $f^{-7/3}$, is the universal consequence of a population of circular binaries whose orbits decay by GW emission with a birth rate approximately constant over the relevant timescales \cite{Phinney:2001di}, and is essentially independent of where in the Galaxy the binaries live. The high-frequency part, controlled by $f_1$, $f_{\rm knee}$ and $\alpha$, is instead sensitive to the population. Assuming $S_h^G$ to be spatially uniform therefore amounts to assuming that (i) the distribution of orbital separations of the unresolved population, equivalently of the ages since the last mass-transfer episode, does not vary appreciably between the bulge, the thin disc and the outer disc, and (ii) the resolvability threshold is direction independent.

Neither statement is exact. The age and metallicity gradients between bulge and disc are expected to modulate both the relative abundance of double white dwarfs per unit stellar mass, their period distribution and the level of observability of resolved sources. For that reason, $\alpha$, $f_1$ and $f_{\rm knee}$ should in principle inherit a weak angular dependence as well. Relaxing either assumption requires replacing the smooth density profile of Eq.~\eqref{eq:rho_galaxy} with a population-synthesis catalogue carrying a position-dependent formation history, which would provide a more accurate mapping between stellar populations and GW sources; we leave this to future work.

\section{Forecast}
\label{sec:forecast}
Following \cite{LISACosmologyWorkingGroup:2022kbp,Mentasti:2023uyi}, we forecast the detectability of the deviation of the model that takes into account the Galaxy rotation from the rotation-free one.
In particular, we define $\theta=(A,f_1,f_2,f_{\rm knee},\alpha)$ the set of free parameters describing the frequency shape of the Galactic model (see Eq. \eqref{eq:PSD_galaxy}) and $\bar\theta$ their true value. The effect of the background on the correlator between instrument $i$ and $j$ of LISA for a given set of parameters at a given time and frequency comes from the angular integral over all the sky directions
\begin{align}
C_{ij}(f,t;\theta)&=\int d^2\hat n\,S_h(f,\hat n;\theta)\mathcal{R}_{ij}(f,\hat n,t)\,,
\end{align}
while the same expected value of the correlator that one would model wrongly without accounting for the rotation effect is
\begin{align}
C_{ij}^0(f,t;\theta)&=S_h^G(f;\theta)\int d^2\hat n\,\mathcal{M}(\hat n)\mathcal{R}_{ij}(f,\hat n,t)\,,
\end{align}
where $\mathcal{R}_{ij}(f,\hat n,t)$ are the LISA response functions defined in a similar way to \cite{LISACosmologyWorkingGroup:2022kbp}:
\begin{align}
&\mathcal{R}_{ij}(f,\hat n,t)\equiv\frac{1}{8\pi}e^{-2\pi if\hat n\cdot(\vec s_i(t)-\vec s_j(t))}\times\nonumber\\
&\sum_A R^A(f\hat n,\hat l_{i,i+1}(t),\hat l_{i,i+2}(t)) R^{A*}(f\hat n,\hat l_{j,j+1}(t),\hat l_{j,j+2}(t))\,.
\end{align}
which encodes the instrumental response to GW coming from the line of sight $\hat{n}$, and depends on frequency $f$, arm unit vector $\hat{l}_{ij}$, and spacecraft positions $\vec{s}_{i}$. The sum runs on the two polarizations $A=+,\times$.
The log-likelihood function in the parameters $\theta$, assuming the correct model (that takes into account the rotation of the galaxy), will be
\begin{align}\label{eq:loglik_true}
\log\mathcal{L}_{\rm r}(\theta)=-\frac{1}{2}\sum_{ij}\int df\int dt \frac{\left|C_{ij}(f,t;\theta)-C_{ij}(f,t;\bar\theta)\right|^2}{D_{ij}(f,t;\bar\theta)}\,,
\end{align}
while using the model that does not take into account the galaxy rotation, one would write
\begin{align}\label{eq:loglik_false}
\log\mathcal{L}_{\rm n}(\theta)=-\frac{1}{2}\sum_{ij}\int df\int dt \frac{\left|C_{ij}^0(f,t;\theta)-C_{ij}(f,t;\bar\theta)\right|^2}{D_{ij}(f,t;\bar\theta)}\,,
\end{align}
with the $D_{ij}(f,t;\bar\theta)$ functions that are defined in an analogous way of \cite{Mentasti:2023uyi}. We conveniently work with the time-delay interferometry (TDI) variables AET, where the noise covariance is diagonal, assuming a simplified equilateral triangle instrumental configuration:
\begin{align}
D_{ij}&(f,t;\bar\theta)=(C_{ii}^{0*}(f,t;\bar\theta)+N_i(f))(C_{jj}^0(f,t;\bar\theta)+N_{j}(f))\nonumber\\
&+(C_{ij}^{0*}(f,t;\bar\theta)+\delta_{ij}N_i(f))(C_{ji}^0(f,t;\bar\theta)+\delta_{ij}N_{j}(f))\,,
\end{align}
with $N_i(f)$ are the noise PSD functions for TDI variable~$i \in \{A, E, T\}$, taken from \cite{LISACosmologyWorkingGroup:2022kbp}. Both $N_i(f)$ and $C_{ij}(f,t)$ can be computed in any TDI combination, provided that the two are consistent with one another. For the sake of our work, we chose the simplest TDI generation 1, but the results do not change for any other combination~\cite{LISACosmologyWorkingGroup:2022kbp,Mentasti:2023uyi}.

\subsection{Bias, error, SNR}
The likelihood function in \eqref{eq:loglik_true} is unbiased by definition: the best fit value $\theta_{\rm fit, r}$ for the galactic background parameters, such that
\begin{align}
\frac{\partial\log\mathcal{L}_{\rm r}(\theta)}{\partial\theta}=0\,,
\end{align}
is $\theta_{\rm fit, r}=\bar\theta$. However, the best fit value obtained by using the wrong model of likelihood \eqref{eq:loglik_false}, such that
\begin{align}
\frac{\partial\log\mathcal{L}_{\rm n}(\theta)}{\partial\theta}=0\,,
\end{align}
is $\theta_{\rm fit, n}\equiv\bar\theta+b_\theta$, where the bias $b_\theta$ is the bias in the estimate of the set of parameters $\theta$.
After some straightforward algebra, one can estimate at first order in $b_\theta/\bar\theta$ the bias \cite{Baghi:2026dfk}
\begin{align}\label{eq:bias}
b_\theta&=\sum_{\theta'}\mathcal{F}_{\theta\theta'}^{-1}\sum_{ij}\int df\int dt\nonumber\\
\times&\Re\left[ \frac{\left(C_{ij}(f,t;\bar\theta)-C_{ij}^0(f,t;\bar\theta)\right)^*\partial_{\theta'} C_{ij}^0(f,t,\bar\theta)}{D_{ij}(f,t;\bar\theta)}\right]\nonumber\\
\mathcal{F}_{\theta\theta'}&=\sum_{ij}\int df\int dt \frac{\Re\left[\partial_\theta C_{ij}^0(f,t,\bar\theta)\partial_{\theta'} C_{ij}^{0*}(f,t,\bar\theta)\right]}{D_{ij}(f,t;\bar\theta)}
\end{align}
and compare it with the forecast error estimated through the Fisher approximation
\begin{align}\label{eq:sigma}
\sigma_\theta=\left(\mathcal{F}^{-1}\right)_{\theta\theta}^{\frac 1 2}
.
\end{align}
We stress that both $b_\theta$ and $\sigma_\theta$ depend on which subset of $\theta$ is treated as free, since $\mathcal{F}$ is inverted only in the corresponding block. Fixing parameters at their true values removes degeneracy directions, which shrinks $\sigma_\theta$ but also prevents the mismatch from being partially absorbed by the remaining freedom. Both cases are reported below.

Another way to quantify the observability of the effect of rotation, which is instead independent of the choice of free parameters, is to compute the signal-to-noise ratio (SNR) of the residual between the correct model and the one that does not take into account the galactic rotation:
\begin{align}\label{eq:snr}
{\rm SNR}_{\rm diff}=\sqrt{\sum_{ij}\int df\int dt \frac{\left|C_{ij}(f,t,\bar\theta)-C_{ij}^0(f,t,\bar\theta)\right|^2}{D_{ij}(f,t,\bar\theta)}}\,.
\end{align}

\subsection{Other detector networks and noise budgets}
\label{sec:networks}

Because the effect turns out to be small for LISA alone, we consider two extensions of the
baseline configuration, both of which act on the same expressions above through the index
$i$ and through $N_i(f)$ only.

The first is the addition of a second constellation, Taiji \cite{Hu:2017mde,Ruan:2018tsw},
which is planned to fly in a heliocentric orbit leading the Earth by $\simeq20^\circ$, while
LISA trails it by a comparable angle. Following \cite{Mentasti:2023uyi}, we model it as an equilateral triangle of arm length
$L_{\rm T}=3\times10^9$ m, ahead of LISA by $40^\circ$ of orbital phase and with its own
constellation orientation, and we build its $A,E,T$ combinations in the same way, with the
noise budget of \cite{Hu:2017mde}. The index $i$ then runs over six channels rather than
three, the three LISA ones and the three Taiji ones, and the sums in
Eqs.~\eqref{eq:loglik_true}--\eqref{eq:snr} run over all $6\times6$ pairs.

The second is a uniform reduction of the instrumental noise. We divide every
$N_i(f)$ by a factor $\mathcal{N}$, keeping the Galaxy model, the response and the
fiducial parameters fixed, and quote results for $\mathcal{N}=1$, i.e.\ the nominal
LISA and Taiji noise budgets of \cite{LISACosmologyWorkingGroup:2022kbp}, and for
$\mathcal{N}=100$, i.e.\ a factor $10$ better noise \emph{amplitude}. The specific choice $\mathcal{N}=100$ is not arbitrary: gains
of this order are contemplated by concrete post-LISA concepts. The LISAmax proposal
\cite{Martens:2023mrc}, a triangular constellation at the Sun--Earth $L_3$, $L_4$ and $L_5$
points with an arm length of $2.59\times10^{8}$ km, is designed to improve the amplitude
sensitivity by at least two orders of magnitude with respect to LISA. We stress, however,
that our $\mathcal{N}=100$ case is a controlled sensitivity study and not a forecast for
that or any other specific mission: it rescales the noise while holding the constellation
geometry, and hence the sky response, fixed, whereas LISAmax lengthens the arm by a factor
$\sim10^2$. Consequently, both shifts its band of best sensitivity below $1$ mHz and
changes the angular response entirely: a forecast for a specific successor would require
recomputing $\mathcal{R}_{ij}(f,\hat n,t)$ for its own configuration, which we leave to
future work.

\subsection{Results of the forecast}
\label{sec:results}
With the nominal noise budget we find ${\rm SNR}_{\rm diff}(2\,{\rm yr})\simeq0.20$ and
${\rm SNR}_{\rm diff}(5\,{\rm yr})\simeq0.22$ for LISA alone, and $0.72$ and $0.79$ for the
LISA+Taiji network\footnote{For reference, the absolute values of the total SNR of the
foreground are about $1044$ and $1119$ for LISA at 2 and 5 years, and $2106$ and $2363$ for
the network.}. The residual between the two templates is therefore \emph{not} detectable
by LISA, nor by the LISA+Taiji network, at the nominal sensitivity. Note also that
${\rm SNR}_{\rm diff}$ does not grow with the square root of the observation time, because
the GW background spectrum in Eq.~\eqref{eq:PSD_galaxy} decreases with time: the longer the
observation, the more GBs are resolved and therefore removed from the data stream.

The ratio ${\rm SNR}_{\rm diff}/{\rm SNR}_{\rm tot}\simeq1.9\times10^{-4}$ for LISA after two
years of observation provides a compact, normalization-independent measure of the size of the
effect. It is much smaller than the naive estimate obtained by multiplying the peak Doppler
shift $|v_{\rm los}|/c\simeq2.8\times10^{-4}$ by the logarithmic slope
$|d\log S^G_h/d\log f|\simeq20$ reached at the top of the band, which would give
$\simeq6\times10^{-3}$. The reason is the near-cancellation already visible in
Figure~\ref{fig:map}: what a detector whose response covers a large fraction of the sky at
once measures is not the peak of $\langle v_{\rm los}\rangle$ but essentially its
$\mathcal{M}$-weighted sky average, $\simeq-4.8$ km/s, i.e.\ $|\langle v_{\rm los}\rangle|_\mathcal{M}/c\simeq1.6\times10^{-5}$.
Multiplying that by the same logarithmic slope gives $\simeq3\times10^{-4}$, within a factor
of two of the measured value. The remaining anisotropic power is what LISA is least able to
pick up, since the autocorrelation of a single triangular constellation is invariant under
$\hat n\to-\hat n$ and is therefore blind to the odd multipoles that dominate the residual.
This is precisely why the LISA+Taiji network gains a factor $3.6$ in ${\rm SNR}_{\rm diff}$
while gaining only a factor $2.0$ in ${\rm SNR}_{\rm tot}$.

\begin{figure}[t]
    \centering
    \includegraphics[width=\linewidth]{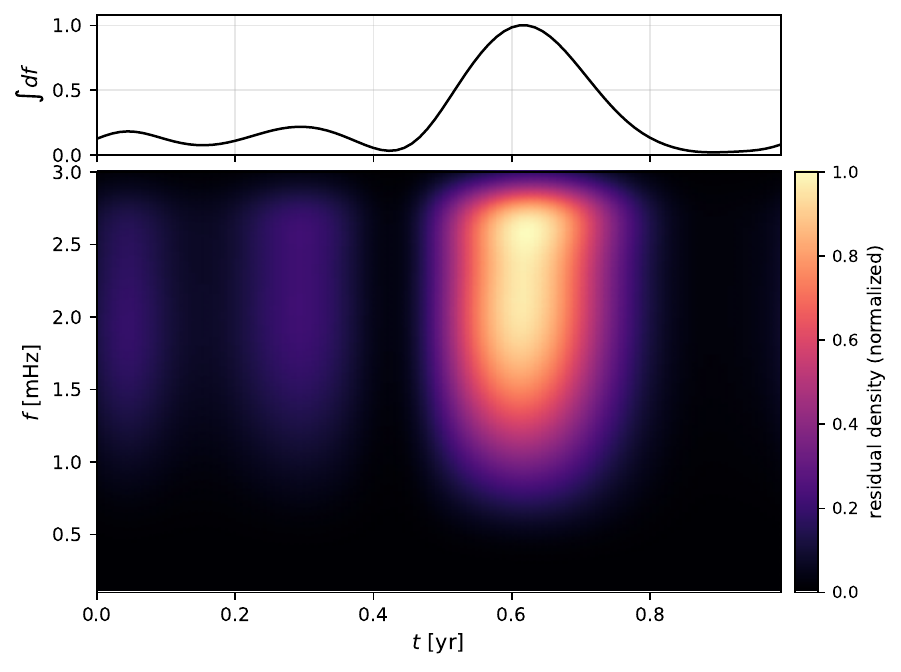}
    \caption{Distribution in the time--frequency plane of the residual between the model that includes the Galaxy rotation and the one that does not, i.e.\ of the integrand of Eq.~\eqref{eq:snr} normalized to its maximum. The upper panel shows the same quantity after integration over frequency. The annual modulation reflects the rotation of the LISA antenna pattern over the velocity map of Figure~\ref{fig:map}.}
    \label{fig:tf}
\end{figure}

Because the effect is by construction a time-dependent one, it is instructive to look at where in the $(f,t)$ plane the residual of Eq.~\eqref{eq:snr} builds up. This is shown in Figure~\ref{fig:tf}, which displays the integrand of Eq.~\eqref{eq:snr}.

We now verify that the forecast bias and error of Eqs.~\eqref{eq:bias} and \eqref{eq:sigma}
well approximate the posterior distribution of the Galactic model spectral parameters
by MCMC sampling with \textit{emcee} \cite{Foreman_Mackey_2013}, using the rotation-blind
log-likelihood of Eq.~\eqref{eq:loglik_false} while the data injection always includes the
rotation effect. We sample the two parameters $(A,f_1)$, which control the overall
amplitude and the high-frequency fall-off of the spectrum, holding the remaining three at
their true values; the bias is then estimated as the difference between the posterior mean
and the injected value, and $\sigma_\theta$ as the standard deviation of the same chain.
For each of the eight configurations we run $32$ walkers for $3000$ steps and discard the
first $1000$ as burn-in, which gives integrated autocorrelation times of $20$--$40$ steps
and $\gtrsim1500$ independent samples per parameter. The comparison is collected in
Table~\ref{tab:fisher_mcmc}.

\begin{table*}[t]
\centering
\footnotesize
\setlength{\tabcolsep}{4.5pt}
\renewcommand{\arraystretch}{1.15}
\begin{tabular}{l c c c | c c c c | c c c c}
\hline\hline
 & & & & \multicolumn{4}{c|}{\textbf{Amplitude} $A$} & \multicolumn{4}{c}{\textbf{Cutoff frequency} $f_1$} \\
 & & & & \multicolumn{2}{c}{$\sigma_A/\bar A$ [\%]} & \multicolumn{2}{c|}{$b_A/\sigma_A$} & \multicolumn{2}{c}{$\sigma_{f_1}/\bar f_1$ [\%]} & \multicolumn{2}{c}{$b_{f_1}/\sigma_{f_1}$} \\
Network & $T_{\rm obs}$ & $\mathcal{N}$ & ${\rm SNR}_{\rm diff}$ & Fisher & MCMC & Fisher & MCMC & Fisher & MCMC & Fisher & MCMC \\
\hline
LISA & 2 yr & 1 & 0.198 & 0.199 & 0.197 & -0.018 & -0.026\,(0.020) & 0.096 & 0.095 & -0.029 & -0.006\,(0.020) \\
LISA & 5 yr & 1 & 0.221 & 0.193 & 0.194 & -0.017 & +0.022\,(0.020) & 0.091 & 0.093 & -0.034 & -0.069\,(0.020) \\
LISA+Taiji & 2 yr & 1 & 0.718 & 0.092 & 0.092 & -0.067 & -0.092\,(0.021) & 0.045 & 0.045 & -0.046 & -0.015\,(0.021) \\
LISA+Taiji & 5 yr & 1 & 0.791 & 0.086 & 0.085 & -0.067 & -0.031\,(0.020) & 0.042 & 0.042 & -0.068 & -0.119\,(0.021) \\
\hline
LISA & 2 yr & 100 & 0.976 & 0.045 & 0.045 & -0.046 & -0.032\,(0.023) & 0.018 & 0.018 & -0.178 & -0.180\,(0.022) \\
LISA & 5 yr & 100 & 1.554 & 0.032 & 0.032 & -0.033 & -0.042\,(0.021) & 0.012 & 0.012 & -0.333 & -0.339\,(0.021) \\
LISA+Taiji & 2 yr & 100 & 2.779 & 0.025 & 0.025 & -0.186 & -0.151\,(0.023) & 0.011 & 0.011 & -0.217 & -0.215\,(0.022) \\
LISA+Taiji & 5 yr & 100 & 4.644 & 0.017 & 0.017 & -0.212 & -0.214\,(0.021) & 0.007 & 0.007 & -0.507 & -0.519\,(0.021) \\
\hline\hline
\end{tabular}
\caption{Forecast statistical uncertainty $\sigma_\theta$ and systematic bias $b_\theta$
induced by neglecting the Galactic rotation, for the inference in which only $(A,f_1)$ are
free and the remaining three spectral parameters of Eq.~\eqref{eq:PSD_galaxy} are held at
their true values. Results are given for LISA and for the LISA+Taiji network, for
$T_{\rm obs}=2$ and $5$ yr, and for two noise levels: $\mathcal{N}=1$ is the nominal noise
budget of \cite{LISACosmologyWorkingGroup:2022kbp} and $\mathcal{N}=100$ divides every noise
PSD by $100$, i.e.\ improves the noise amplitude by a factor $10$ (Section \ref{sec:networks}).
Uncertainties are quoted relative to the fiducial value $\bar\theta$, biases relative to
$\sigma_\theta$. ``Fisher'' refers to Eqs.~\eqref{eq:bias} and \eqref{eq:sigma}. The fiducial values are $\bar A=1.259\times10^{-44}$ for both durations, and
$\bar f_1=1.678$ mHz ($1.334$ mHz) for $T_{\rm obs}=2$ yr ($5$ yr).}
\label{tab:fisher_mcmc}
\end{table*}

\begin{table*}[t]
\centering
\footnotesize
\setlength{\tabcolsep}{3.4pt}
\renewcommand{\arraystretch}{1.15}
\begin{tabular}{l c c | c c c c c | c c c c c}
\hline\hline
 & & & \multicolumn{5}{c|}{\textbf{Marginal error} $\sigma_\theta/\bar\theta$ [\%]} & \multicolumn{5}{c}{\textbf{Bias} $b_\theta/\sigma_\theta$} \\
Network & $T_{\rm obs}$ & $\mathcal{N}$ & $A$ & $f_1$ & $f_2$ & $f_{\rm knee}$ & $\alpha$ & $A$ & $f_1$ & $f_2$ & $f_{\rm knee}$ & $\alpha$ \\
\hline
LISA & 2 yr & 1 & 0.711 & 0.432 & 2.389 & 0.170 & 0.991 & -0.008 & -0.003 & -0.001 & -0.010 & +0.001 \\
LISA & 5 yr & 1 & 0.787 & 0.460 & 1.512 & 0.245 & 1.131 & -0.008 & -0.003 & -0.001 & -0.007 & +0.001 \\
LISA+Taiji & 2 yr & 1 & 0.310 & 0.191 & 1.085 & 0.081 & 0.452 & -0.019 & -0.010 & +0.002 & -0.028 & -0.007 \\
LISA+Taiji & 5 yr & 1 & 0.331 & 0.195 & 0.687 & 0.113 & 0.496 & -0.014 & -0.017 & -0.004 & -0.018 & -0.007 \\
\hline
LISA & 2 yr & 100 & 0.104 & 0.071 & 0.298 & 0.033 & 0.176 & -0.044 & -0.015 & -0.007 & -0.038 & +0.003 \\
LISA & 5 yr & 100 & 0.080 & 0.054 & 0.074 & 0.027 & 0.128 & -0.062 & -0.017 & -0.046 & -0.038 & +0.010 \\
LISA+Taiji & 2 yr & 100 & 0.054 & 0.038 & 0.185 & 0.020 & 0.099 & -0.083 & -0.050 & +0.008 & -0.090 & -0.038 \\
LISA+Taiji & 5 yr & 100 & 0.040 & 0.028 & 0.041 & 0.015 & 0.069 & -0.103 & -0.099 & -0.123 & -0.076 & -0.044 \\
\hline\hline
\end{tabular}
\caption{Marginal statistical errors and bias-to-error ratios for the inference in which
all five spectral parameters of Eq.~\eqref{eq:PSD_galaxy} are left free, computed from the
Fisher expressions \eqref{eq:bias} and \eqref{eq:sigma}, for the same eight configurations
as in Table~\ref{tab:fisher_mcmc}. Errors are quoted relative to the fiducial value, whose
full set is $\bar A=1.259\times10^{-44}$, $\bar f_2=0.316$ mHz and $\bar\alpha=1.80$ for
both durations, with $\bar f_1=1.678$ mHz and $\bar f_{\rm knee}=2.810$ mHz for
$T_{\rm obs}=2$ yr and $\bar f_1=1.334$ mHz, $\bar f_{\rm knee}=2.194$ mHz for
$T_{\rm obs}=5$ yr.}
\label{tab:fisher_5par}
\end{table*}

Tables~\ref{tab:fisher_mcmc} and \ref{tab:fisher_5par} tell a consistent story. At the
nominal noise level the systematic error induced by neglecting the Galactic rotation is
negligible: the bias-to-error ratio never exceeds the unity, for either instrument configuration, at either
observation time, and in either the two- or the five-parameter inference. The Fisher and MCMC estimates agree on $\sigma_\theta$ to better than $2.5\%$
in all eight configurations, whereas the bias ratios scatter by a few hundredths and, in the
rows where the bias is smallest, even differ in sign.
S
The reason is that the two quantities are extracted from different moments of the same
chain. With $N_{\rm eff}$ effective samples, the standard deviation is determined to a
fractional accuracy $1/\sqrt{2N_{\rm eff}}\simeq1.4\%$, the level at which $\sigma_\theta$
agrees. On the other hand, the posterior mean is determined only to $\sigma_\theta/\sqrt{N_{\rm eff}}$,
i.e.\ to $\simeq0.02$ in the units of $\sigma_\theta$ in which the bias is tabulated (the
number in parentheses in Table~\ref{tab:fisher_mcmc}). At nominal noise, where
$|b_\theta/\sigma_\theta|\lesssim0.07$, the MCMC therefore measures the bias at a
signal-to-noise ratio of order unity, and a sign flip is the expected outcome rather than a
discrepancy. We have checked this in two ways: locating the minimum of the rotation-blind
likelihood \eqref{eq:loglik_false} numerically reproduces the Fisher bias to three decimal
places in all configurations. In this way, the first-order expression \eqref{eq:bias} is exact at the
precision we quote; and chains repeated from independent seeds scatter around the Fisher
values with exactly the predicted width.

The smallness of the bias is not independent of the smallness of ${\rm SNR}_{\rm diff}$. We write
$\langle x,y\rangle\equiv\sum_{ij}\int df\int dt\,\Re[x_{ij}^*y_{ij}]/D_{ij}$ for the inner product that
appears in Eqs.~\eqref{eq:bias}--\eqref{eq:snr}, so that $\mathcal{F}_{\theta\theta'}=\langle\partial_\theta C^0,\partial_{\theta'}C^0\rangle$
and ${\rm SNR}_{\rm diff}^2=\langle\Delta,\Delta\rangle$ with $\Delta\equiv C-C^0$. Two distinct inner products enter here, and it is worth keeping them apart.
The first, $\langle\cdot,\cdot\rangle$, acts on functions of frequency and time, i.e.\ on
correlators; it is the one that defines both $\mathcal{F}$ and ${\rm SNR}_{\rm diff}$. The
second acts on vectors in the space of the free parameters and has metric
$\mathcal{F}^{-1}$; the Fisher matrix is precisely the metric that the first inner product
induces on parameter space through the derivatives $\partial_\theta C^0$. The bound below
follows from applying the Cauchy--Schwarz inequality once in each: in parameter space, with
metric $\mathcal{F}^{-1}$, to $b_\theta=(\mathcal{F}^{-1}n)_\theta$ and
$\sigma_\theta^2=(\mathcal{F}^{-1})_{\theta\theta}$, which gives the first inequality, and
then in correlator space, to $n_\theta=\langle\Delta,\partial_\theta C^0\rangle$, whose squared norm
$n^{T}\mathcal{F}^{-1}n$ is the squared length of the projection of $\Delta$ onto the subspace
spanned by the template derivatives and therefore cannot exceed
$\langle\Delta,\Delta\rangle$, which gives the second. The derivation is given in full in Appendix~\ref{sec:bound}, together with a discussion of the meaning of the two inner products. Altogether
\begin{align}\label{eq:bound}
\left|\frac{b_\theta}{\sigma_\theta}\right|\le\sqrt{n^{T}\mathcal{F}^{-1}n}\;\le\;{\rm SNR}_{\rm diff}\,,
\end{align}
the first inequality being saturated
when $\Delta$ has no overlap with the template derivatives of the other free parameters, $n_{\theta'}=0$ for $\theta'\neq\theta$, and the second when $\Delta$ lies
entirely inside the space spanned by the template derivatives. We report the results numerically: the displacement between the two posteriors is precisely
$b_\theta/\sigma_\theta$ in units of the contour width. In this way, at the values of
Table~\ref{tab:fisher_mcmc} the contours obtained with the two likelihoods differ by a few
per cent of their own size, which is below both the plotted line width and the Monte-Carlo
scatter of the contours themselves.

The dependence on the noise level shows what would have to change for the effect to become
relevant. Dividing every noise PSD by $100$ increases ${\rm SNR}_{\rm diff}$ by a factor
between $3.9$ and $7.0$ --- not by the factor $10$ that a noise-dominated measurement would
give, because the foreground itself begins to dominate the instrumental noise and the
residual saturates. The bias ratios grow correspondingly, reaching $0.33\sigma$ on $f_1$ for
LISA at five years and $0.51\sigma$ for the network, still below the statistical uncertainty
in every case. It is instructive to push this to its limit: setting the instrumental noise to
zero, so that $D_{ij}$ retains only the signal terms, gives an absolute ceiling
\begin{align}
{\rm SNR}_{\rm diff}^{\rm max}=1.08\ (2.95)\quad&\text{for LISA},\nonumber\\
{\rm SNR}_{\rm diff}^{\rm max}=2.98\ (8.14)\quad&\text{for LISA+Taiji},
\end{align}
at $T_{\rm obs}=2$ ($5$) yr. Even an ideal, noise-free instrument of the assumed geometry
could therefore not detect the effect with LISA alone over a nominal two-year mission: the
limitation is not only instrumental but intrinsic, set by the variance of the foreground
against which the residual has to be measured. The $\mathcal{N}=100$ column already realizes
$90\%$ ($53\%$) of the LISA ceiling, and $93\%$ ($57\%$) of the network one, at two (five)
years.

Finally, the pattern of the biases across parameters is the one expected for a coherent
frequency shift. In Table~\ref{tab:fisher_5par} the entries are of the same sign for all the
parameters controlling the position of the cutoff, $A$, $f_1$ and $f_{\rm knee}$, and the
largest of them is on $f_{\rm knee}$ or on $A$ in seven of the eight configurations; the
spectral index $\alpha$, which controls the sharpness rather than the position of the
fall-off, is the least affected parameter in five of them and the second least in the rest. Releasing
the three cutoff-shape parameters inflates the marginal errors, by a factor of $3$ to $4$
because of the degeneracies between $A$, $f_1$ and $\alpha$, and lets part of the mismatch be
absorbed by the extra freedom, so the five-parameter ratios are smaller than their
two-parameter counterparts in every configuration.

\section{Implications for foreground modelling}
\label{sec:implications}

For an individually resolved binary, the Doppler factor of Eq.~\eqref{eq:doppler} shifts the measured frequency $f_0$, but the shift is not observable, because it is completely degenerate with the intrinsic parameters of the binary, whose source-frame frequency is not known a priori~\cite{Ebadi:2024mvn}. The signal of a binary moving with constant velocity is identical to that of a binary at rest with frequency $f_0$ and with chirp mass and distance rescaled by the Doppler factor, exactly as for the cosmological redshift of extragalactic sources~\cite{Bonvin:2016qxr,Tamanini:2019usx}. In the unresolved foreground, instead, the shift varies across the sky and cannot be fully absorbed into the parameters of the factorized template of Eq.~\eqref{spectrum0}; this is the effect quantified in Section~\ref{sec:forecast}.

Other kinematic effects are negligible for the foreground. The acceleration of a binary in the Galactic potential, $a\sim2\times10^{-10}\,{\rm m\,s^{-2}}$ \cite{Silverwood:2018qra}, and the apparent acceleration produced by its transverse motion change the observed frequency only through the variation of the velocity over the mission, by $\Delta f/f\sim a\,T_{\rm obs}/c\sim4\times10^{-11}$ and $\Delta f/f\sim v_\perp^2T_{\rm obs}/(c\,d)\sim3\times10^{-10}$ respectively, for $T_{\rm obs}=2$ yr, $v_\perp\sim200$ km/s and $d\sim1$ kpc, i.e.\ six to seven orders of magnitude below the Doppler shift. They can instead matter for individually resolved binaries, whose frequency evolution is tracked coherently over the whole mission~\cite{Ebadi:2024mvn}.

The pipelines currently used to characterize the foreground, such as the iterative subtraction of Ref.~\cite{Karnesis:2021tsh}, need to be modified only in their last stage. The identification and removal of the resolvable binaries, and the estimate of the confusion level that accompanies it, operate on individual frequency bands, where a distortion of $\sim2\times10^{-4}$ once averaged over the sky and the year is far below the scatter of a single periodogram: they are unaffected, and a test on a single simulated realization is expected to show no change. The effect enters only the final fit of the smooth template to the accumulated time-frequency data, as the systematic quantified in Table~\ref{tab:fisher_mcmc}, and it can be included there at no cost in dimensionality, by replacing $S_h^G(f)\mathcal{M}(\hat n)$ with Eq.~\eqref{eq:Sh_full} in the template correlator, since the Galactic velocity field is fixed by electromagnetic observations.

\section{Conclusions}
\label{sec:conclusions}

In this work, we have investigated the impact of the Milky Way rotation on the stochastic gravitational-wave background (SGWB) as observed by LISA. By combining a realistic model of the Galactic stellar density with a kinematic description based on the observed rotation curve, we have derived the anisotropic GW spectrum including the Doppler modulation induced by the relative motion between sources and the observer.

A key result of our analysis is that the Doppler effect is intrinsically direction-dependent: each line of sight probes a different distribution of velocities, and therefore a different frequency shift of the emitted signal. This leads to distortions of the SGWB that cannot be captured by simpler models assuming a single effective boost. As a consequence, the commonly used factorized description of the signal into a purely spectral term times an angular template breaks down once Galactic kinematics are properly included. Averaged over the sky the effect nearly cancels (Figure~\ref{fig:map}), and the residual signal arises only because the detector response weights the sky unevenly and rotates over the year (Figure~\ref{fig:tf}).

Using both a Fisher matrix formalism and an MCMC analysis, we quantified the impact of neglecting this effect in parameter estimation, for LISA and for the LISA+Taiji network, for two- and five-year missions and for different noise levels. At the nominal sensitivities the effect is below the detection threshold, and the bias it induces on the spectral parameters is well below the statistical uncertainty, with the Fisher and MCMC estimates in agreement. We also showed that the bias, in units of the statistical error, is always bounded by the signal-to-noise ratio of the residual. Neglecting the Galactic rotation is therefore not a limiting systematic, neither for LISA nor for LISA+Taiji.

The effect is much smaller than a naive estimate would suggest, because the line-of-sight velocity largely cancels over the sky, and what survives is an anisotropic pattern whose odd multipoles are invisible to a single triangular detector, though not to the cross-correlation of two. The limitation is partly intrinsic, since even without instrumental noise the variance of the foreground itself caps the significance of the effect. Still, a noise amplitude lower by an order of magnitude would realize most of this ceiling, bringing the effect to the threshold of detection for LISA and above it for LISA+Taiji, with biases that become a sizeable fraction of the statistical error. Post-LISA concepts such as LISAmax \cite{Martens:2023mrc} aim at improvements of this size, though with a very different response, for which a dedicated forecast remains to be done.

Including the effect requires no additional free parameters, only a perfectly known change in the template of the power spectrum. At the nominal sensitivities it does not affect component separation, and in existing foreground-subtraction pipelines it would only enter the final template fit (Section~\ref{sec:implications}).

There are several directions for future work. First, our model assumes a smooth and axisymmetric Galaxy, while real structures such as spiral arms, bars, and local substructures may introduce additional anisotropies and velocity features. Including these ingredients would refine the predictions and possibly enhance the observable signatures. Second, we assumed that the GW emission traces the stellar density with a spatially uniform spectral shape (Section~\ref{sec:assumption}); a population synthesis model would provide a more accurate mapping between stellar populations and GW sources.
Furthermore, extending the analysis to different detector configurations and data analysis pipelines would help assess the robustness of our conclusions, in particular when also resolved sources and other GW backgrounds are present in the data stream.

Lastly, since the galactic background is sensitive to the kinematics of the Galaxy, a GW measurement could in principle constrain the rotation curve, and through it the Galactic potential, independently of electromagnetic surveys. This is out of reach for LISA and LISA+Taiji, and would require a considerably quieter instrument.

\begin{acknowledgments}
GM acknowledges support from ANR grant AAPG2024 PRC - GalaxyFIT. We thank Carlo Contaldi for clarifications on the computation of the galactic spectrum, Stanislav Babak, Henri Inchausp\'e, Germano Nardini for precious discussions at the very beginning of the work, Astrid Lamberts and Alice Perego for insights on the interpretation of the GB background parameters. We also thank the anonymous referee for a careful reading of the manuscript and for several suggestions that substantially improved its presentation.
\end{acknowledgments}

\appendix

\section{Galaxy profile, spectrum and geometry}
\label{sec:galaxy}
The stellar density, as a function of galactocentric cartesian coordinates, is well approximated by \cite{Nelemans:2003ha,PhysRevD.86.124032}
\begin{align}\label{eq:rho_galaxy}
\rho(x,y,z) = \rho_0\left[A_\rho e^{-r^2/R^2_b}+(1-A_\rho)e^{-u/R_d}\sech^2(z/Z_d) \right]\,,
\end{align}
where $r^2=x^2+y^2+z^2 \equiv u^2+z^2$, $R_b$ is the characteristic radius of the bulge, and $R_d$ and $Z_d$ are the characteristic radius and scale height of the disc, respectively. $A_\rho$ gives the relative weight of the stellar density in the bulge compared to the disc. The normalization of the stellar density $\rho_0$ is set to unity as we are only interested in the morphology of the signal. We set $A_\rho=0.25$, $R_b=500$ pc, $R_d=2500$ pc, and $Z_d=200$ pc.

We model the galactic background power spectral density (PSD) in its reference frame with the parametrization of \cite{Karnesis:2021tsh}
\begin{align}\label{eq:PSD_galaxy}
&S_h^G(f)= Af^{-7/3}\,e^{-(f/f_1)^\alpha}\left[1+\tanh\left(\frac{f_{\rm knee}-f}{f_2}\right)\right] \;, \nonumber\\
&\log_{10} (f_1/\text{Hz})=a_1\log_{10}(T_{obs}/\text{yr})+b_1 \;, \nonumber\\
&\log_{10} (f_{\rm knee}/\text{Hz})=a_k\log_{10}(T_{obs}/\text{yr})+b_k\,,
\end{align}
with $A=10^{-43.9}$, $\alpha=1.8$, $a_1=-0.25$, $b_1=-2.7$, $a_k=-0.27$, $b_k=-2.47$, $f_2=10^{-3.5}$Hz.

\subsection{Parametrization of the star velocities}

The stars orbit the galaxy with a velocity that depends on the distance $u$ with respect to the galaxy rotation axis, which can be approximated by a smooth fit from data \cite{McGaugh:2018ruf}. We neglect the individual stars' peculiar velocities, as they are one order of magnitude or more below the leading rotational motion and they do not present a directional pattern in the sky

\begin{align}\label{eq:rotcurve}
v(u)&=v_0\left(1-e^{-u/u_0}\right)+v_b\,e^{-\left(\frac{u-u_b}{w_b}\right)^2}\nonumber\\
&\quad-v_d\,\frac{u-u_d}{1+e^{-(u-u_d)/w_d}}\,,\nonumber\\
u&=\sqrt{x^2+y^2}\,,
\end{align}
with $v_0=225$ km/s, $u_0=1.5$ kpc, $v_b=12$ km/s, $u_b=5$ kpc, $w_b=3$ kpc,
$v_d=2.18$ km/s/kpc, $u_d=10$ kpc and $w_d=1.5$ kpc, and $u$ expressed in kpc.
The three terms describe, respectively, the steep rise of the inner rotation curve, the
bump contributed by the bulge and the inner disc near $u\simeq5$ kpc, and the mild
decline of the outer disc. The resulting curve rises to $\simeq220$ km/s by $u=4$ kpc,
peaks at $232$ km/s near $u=6$ kpc and declines slowly outwards, giving $229$ km/s at the
solar radius, in agreement with the fit of \cite{McGaugh:2018ruf}.

\begin{figure}[t]
    \centering
    \includegraphics[width=0.85\linewidth]{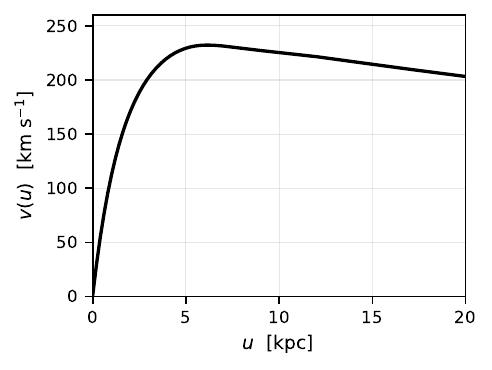}
    \caption{The Galactic rotation curve of Eq.~\eqref{eq:rotcurve} used throughout this
    work. The circular speed rises steeply inside $\simeq3$ kpc, peaks at $232$ km/s near
    $u=6$ kpc, and declines slowly outwards, giving $229$ km/s at the solar radius
    $u_\odot=8$ kpc.}
    \label{fig:vrot}
\end{figure}

Let's place, for the sake of simplicity, the Sun at location $\vec x_s=(x_0,0,0)\,$ with $x_0=8000$ pc and with velocity $\vec v_s=(-11,-240,7)$km/s (this is done in the reference frame where the galactic plane rotates clockwise if seen from the galactic North Pole).
Let $r$ be the physical distance between the observer and a star in the Galaxy.
Then the location $\vec x'$ in the reference frame of the observer as a function of $r$ and $\hat n$ is
\begin{align}
x'(r,\hat n)&=r\sin\theta\cos\phi\,,\nonumber\\
y'(r,\hat n)&=r\sin\theta\sin\phi\,,\nonumber\\
z'(r,\hat n)&=r\cos\theta\,,
\end{align}
and the location in the reference frame of the galaxy $\vec x$ is obtained by a translation $\vec x=\vec x'+\vec x_s$.
Finally, the velocity of the stars at location $\vec x$ is obtained by modeling the motion of each star, located at distance $u$ from the rotation axis, as approximately uniform circular. The (clockwise) rotation of each star is on a circle of radius $u$, with velocity vector of constant norm $v(u)$
\begin{align}
v_x&=v(u)\frac{y}{u}\,,\nonumber\\
v_y&=-v(u)\frac{x}{u}\,,\nonumber\\
v_z&=0\,.
\end{align}
One can easily verify that $v_x^2+v_y^2=v(u)^2$. The velocity of a source relative to the observer is then $\vec v(\vec x)=\vec v_{\rm star}(\vec x)-\vec v_s$.

\section{Derivation of the bound on the bias}
\label{sec:bound}

We derive here Eq.~\eqref{eq:bound} and make explicit the two inner products on which it rests. All derivatives are evaluated at the true parameters $\bar\theta$, and $D_{ij}$ stands for $D_{ij}(f,t;\bar\theta)$.

\subsection{The inner product on correlators}

The inner product that appears in Eqs.~\eqref{eq:bias}--\eqref{eq:snr} is
\begin{align}
\langle x,y\rangle\equiv\sum_{ij}\int df\int dt\,\frac{\Re\left[x_{ij}^*(f,t)\,y_{ij}(f,t)\right]}{D_{ij}(f,t)}\,,
\end{align}
with norm $\|x\|\equiv\langle x,x\rangle^{1/2}$, where $x$ and $y$ stand for correlators or for their differences and derivatives. Since $D_{ij}>0$, the Cauchy--Schwarz inequality $|\langle x,y\rangle|\le\|x\|\,\|y\|$ holds. Here $D_{ij}(f,t)$ is the variance of the measured correlator $C_{ij}$ in the time--frequency cell around $(f,t)$, including both the instrumental noise and the intrinsic variance of the foreground, and $df\,dt$ counts the independent cells. The inner product is therefore the inverse-variance-weighted overlap of two correlator templates, and $\|x\|$ is the signal-to-noise ratio with which a change $x$ of the expected correlators can be detected. In this notation, the likelihoods \eqref{eq:loglik_true} and \eqref{eq:loglik_false} read
\begin{align}
\log\mathcal{L}_{\rm r}(\theta)&=-\tfrac12\,\big\|C(\theta)-C(\bar\theta)\big\|^2\,,\nonumber\\
\log\mathcal{L}_{\rm n}(\theta)&=-\tfrac12\,\big\|C^0(\theta)-C(\bar\theta)\big\|^2\,,
\end{align}
and, with $\Delta\equiv C(\bar\theta)-C^0(\bar\theta)$,
\begin{align}
\mathcal{F}_{\theta\theta'}&=\langle\partial_\theta C^0,\partial_{\theta'}C^0\rangle\,,\nonumber\\
n_\theta&\equiv\langle\Delta,\partial_\theta C^0\rangle=\partial_\theta\log\mathcal{L}_{\rm n}(\bar\theta)\,,\nonumber\\
{\rm SNR}_{\rm diff}&=\|\Delta\|\,,
\end{align}
while Eqs.~\eqref{eq:bias} and \eqref{eq:sigma} read $b=\mathcal{F}^{-1}n$ and $\sigma_\theta^2=(\mathcal{F}^{-1})_{\theta\theta}$.

\subsection{The bias}

At first order in the parameter shift $b$,
\begin{align}
-2\log\mathcal{L}_{\rm n}(\bar\theta+b)\simeq\Big\|\Delta-\sum_\theta b_\theta\,\partial_\theta C^0\Big\|^2\,,
\end{align}
which is minimized when, for every $\theta$,
\begin{align}\label{eq:normal}
0&=\Big\langle\partial_\theta C^0,\,\Delta-\sum_{\theta'}b_{\theta'}\,\partial_{\theta'}C^0\Big\rangle\nonumber\\
&=n_\theta-\sum_{\theta'}\mathcal{F}_{\theta\theta'}\,b_{\theta'}\,,
\end{align}
i.e.\ for $b=\mathcal{F}^{-1}n$, which is Eq.~\eqref{eq:bias}. Let $\Delta_\parallel\equiv\sum_\theta b_\theta\,\partial_\theta C^0$ be the change of the template produced by the bias, and $\Delta_\perp\equiv\Delta-\Delta_\parallel$. By Eq.~\eqref{eq:normal}, $\Delta_\perp$ is orthogonal to every $\partial_\theta C^0$, and hence to $\Delta_\parallel$, which is therefore the projection of the residual onto the template derivatives: $\Delta_\parallel$ is the part of the rotation effect that the rotation-blind template absorbs by shifting its parameters, and $\Delta_\perp$ is left over as the misfit, $-2\log\mathcal{L}_{\rm n}(\bar\theta+b)=\|\Delta_\perp\|^2$. Moreover,
\begin{align}\label{eq:Dpar}
\|\Delta_\parallel\|^2&=\langle\Delta_\parallel,\Delta\rangle=\sum_\theta b_\theta\,n_\theta\nonumber\\
&=n^T\mathcal{F}^{-1}n=b^T\mathcal{F}\,b\,.
\end{align}

\subsection{The two inequalities}

The second inner product acts on vectors of parameter space,
\begin{align}
(u,v)\equiv u^T\mathcal{F}^{-1}v\,.
\end{align}
It is the one that $\langle\cdot,\cdot\rangle$ induces on gradients such as $n$: $\mathcal{F}^{-1}$ turns $n$ into the parameter shift $b$, which changes the template by $\Delta_\parallel$, and $(n,n)=\|\Delta_\parallel\|^2$ by Eq.~\eqref{eq:Dpar}. With $e_\theta$ the unit vector along the parameter $\theta$, one also has $b_\theta=(e_\theta,n)$ and $\sigma_\theta^2=(e_\theta,e_\theta)$, so that the Cauchy--Schwarz inequality for $(\cdot,\cdot)$ gives the first inequality of Eq.~\eqref{eq:bound},
\begin{align}
|b_\theta|=|(e_\theta,n)|&\le(e_\theta,e_\theta)^{\frac12}\,(n,n)^{\frac12}\nonumber\\
&=\sigma_\theta\,\sqrt{n^T\mathcal{F}^{-1}n}\,,
\end{align}
saturated when $n$ is parallel to $e_\theta$, i.e.\ when $n_{\theta'}=0$ for every $\theta'\neq\theta$. The Cauchy--Schwarz inequality for $\langle\cdot,\cdot\rangle$, applied to Eq.~\eqref{eq:Dpar}, gives the second,
\begin{align}
n^T\mathcal{F}^{-1}n=\|\Delta_\parallel\|^2=\langle\Delta_\parallel,\Delta\rangle\le\|\Delta_\parallel\|\,\|\Delta\|\,,
\end{align}
i.e.\ $\sqrt{n^T\mathcal{F}^{-1}n}=\|\Delta_\parallel\|\le\|\Delta\|={\rm SNR}_{\rm diff}$, saturated when $\Delta_\perp=0$, that is when the whole effect is absorbed into the bias. Equivalently, since $\Delta_\parallel$ and $\Delta_\perp$ are orthogonal,
\begin{align}
{\rm SNR}_{\rm diff}^2=b^T\mathcal{F}\,b+\|\Delta_\perp\|^2\,:
\end{align}
the squared signal-to-noise ratio of the rotation effect is shared between the part absorbed into the bias and the misfit, and neither can exceed the total. Combining the two inequalities gives Eq.~\eqref{eq:bound}. The derivation holds for any choice of free parameters, and the upper bound ${\rm SNR}_{\rm diff}$ is the same for all of them. It relies on the same linearization in $b$ as Eq.~\eqref{eq:bias}, which we verified in Section~\ref{sec:results} against the exact minimum of the rotation-blind likelihood.


\bibliography{refs_v3}

\end{document}